\documentclass[preprints,article,accept,moreauthors,pdftex]{Definitions/mdpi}
\usepackage{lineno}
\usepackage{ulem}
\usepackage{paracol}
\firstpage{1}
\pubvolume{8}
\issuenum{2}
\articlenumber{134}
\pubyear{2022}
\copyrightyear{2022}
\externaleditor{\textls[-15]{Academic Editor: Clementina Agodi}}
\datereceived{29 December 2021}
\dateaccepted{19 February 2022}
\datepublished{21 February 2022}
\hreflink{https://doi.org/ \linebreak10.3390/universe8020134} 
\pdfoutput=1

\Title{Experimental study of the positive ion feedback from gas to liquid in a dual-phase argon chamber and measurement of the ion mobility in argon gas}

\TitleCitation{Title}

\Author{Luciano~Romero $^{1}$, Roberto~Santorelli $^{1}$, Edgar~S\'anchez~Garc\'ia $^{1,\dagger,}$*, Thorsten~Lux $^{2}$, Michael~Leyton $^{2}$, Silvestro~di~Luise $^{2}$, Pablo~Garc\'{\i}a~Abia$^{1}$, Rodrigo~L\'opez~Manzano$^{1}$, Jos\'e~Manuel~Cela-Ruiz$^{1}$, Sebasti\'an~Quizhpi$^{1}$ and Vicente~Pesudo$^{1}$. }

\AuthorNames{Firstname Lastname, Firstname Lastname and Firstname Lastname}

\AuthorCitation{Lastname, F.; Lastname, F.; Lastname, F.}

\address{%
$^{1}$ \quad Centro de Investigaciones Energ\'{e}ticas, Medioambientales y Tecnol\'{o}gicas (CIEMAT), Av.~Complutense~40, 28040 Madrid, Spain \\
$^{2}$ \quad Institut de F\'{i}sica d\'{}Altes Energies (IFAE) -  The Barcelona Institute of Science and Technology (BIST), Campus UAB, 08193 Bellaterra (Barcelona), Spain \\ 
$^{3}$ \quad Universidad Complutense de Madrid, Spain }

\firstnote{Current address: Max-Planck-Institut f\"ur Kernphysik, Saupfercheckweg 1, 69117 Heidelberg, Germany} 

\corres{Correspondence: esanchez@mpi-hd.mpg.de}

\abstract{The dynamics of the positive ions created by particle interactions inside  argon time projection chambers plays an important role in characterizing the next generation of massive detectors planned for the direct search for dark matter and the study of neutrino properties. We have constructed a 1~L liquid argon chamber (ARION: ARgon ION experiment) with a high voltage pulse generator capable of injecting, in a controlled manner, a sizeable ion current into the drift region.  This chamber is capable of reproducing a volume charge similar to that found in large detectors, allowing its effects to be studied systematically.  New experimental results regarding ion dynamics in the liquid and direct demonstration of ion feedback from the gas to the liquid are discussed in this paper. In addition, a novel technique to measure the drift velocity of argon ions is introduced along with preliminary results obtained in gas.}


\begin{document}


\section{Introduction}
\label{sec:intro}

Liquid argon (LAr) Time Projection Chambers (TPC) are playing an increasingly important role in dark matter (DM) ~\cite{DS50:2018ves,DEAP:2019imk,DS20:2017fik} and neutrino experiments \cite{uBOONE:2016smi,DUNE}. LAr has the advantage of allowing the construction of large-mass detectors at a reasonable cost, while also enabling the implementation of a fully active tracking calorimeter.

In a typical LAr-TPC,  photon sensors detect prompt scintillation light, while electrons drift to the anode under a uniform electric field $\vec{E_{d}}$. The drift field suppresses the prompt recombination of oppositely charged particles created along the ionization track, causing them to drift towards the anode or cathode along the electric field lines. The mobility, $\mu_i$, of argon ions (Ar$^{+}_{2}$) in liquid is much smaller than that of electrons ($\upmu_{i}  \ll \upmu_{e}$); the latter are thus promptly collected while the former stay in the drift region for a much longer time. For a field of $E_d = 1$~kV/cm, the electron velocity is of the order of $v_e \approx 2$~mm/$\mu$s \cite{Walkowiak:2000wf}. The ion mobility is not well known and the values reported in the literature range between approximately $2\cdot 10^{-4}~\text{cm}^2\,\text{V}^{-1}\text{s}^{-1}$ \cite{Dey:1968} and $1.6\cdot 10^{-3}~\text{cm}^2\,\text{V}^{-1}\text{s}^{-1}$~\cite{ICARUS:2015torti,atlas} with the liquid in a steady state. Considering the larger value, the expected ion velocity at 1~kV/cm is $v_i \approx 1.6\cdot 10^{-5}$~mm/$\mu$s.  
The ions therefore spend substantially more time in the drift region before being neutralized at the cathode. As a consequence, the average density of positive ions in the active volume, $d_{i}$, is much larger than that of electrons, $d_{e}$. The positive charge density in the active volume increases with time, eventually reaching a constant  value depending on the particle interaction rate in the detector. 

This positive space charge can locally modify the drift lines, the magnitude of the electric field, and ultimately the velocity of the  electrons, eventually producing a  displacement in the reconstructed position of the ionisation signal. Additionally, for relatively large values of $d_{i}$, the high positive-charge density makes it necessary to consider the probability of ``secondary recombination'' between drifting electrons and free positive ions. This process is different from the recombination that occurs within the ionisation track \cite{Chepel:2012sj}. The secondary recombination can cause signal loss, with a probability dependent on the electron drift path, comparable to the charge quenching due to contamination of the electronegative impurities in LAr.  At the same time, it can produce the emission of photons through channels similar to the primary recombination light. 

The effects caused by the positive volume charge in massive argon detectors  can be particularly relevant in case of charge amplification at the anode, as in a dual-phase (liquid-gas) TPC. In this case, the ions created in the vapour volume may cross the gas-liquid interface  and further increase  $d_{i}$ in the active LAr volume.

Previous experiments have measured space charge effects in small-scale volumes in liquid argon single-phase detectors~\cite{Palestini:1998an,Rutherfoord:2015mfn}. However, these effects have not been studied with a small dual-phase detector. To study the dynamics of the argon ions in liquid and measure the ion feedback from the gas into the liquid phase, we constructed the ARION (ARgon ION) experiment. The setup consists of a small liquid argon chamber and a system that injects defined positive ion currents in the drift region, producing a sizeable volume charge. In this article, we discuss recent experimental results regarding the dynamics of ions in liquid and the direct measurement of ion feedback from the gas into the liquid phase. Additionally, we present a new measurement of the drift velocity of argon ions in the gas phase.

\section{Experimental setup} 
\end{paracol}

\begin{figure}[!t]
\centering
\includegraphics[width=0.4\textwidth]{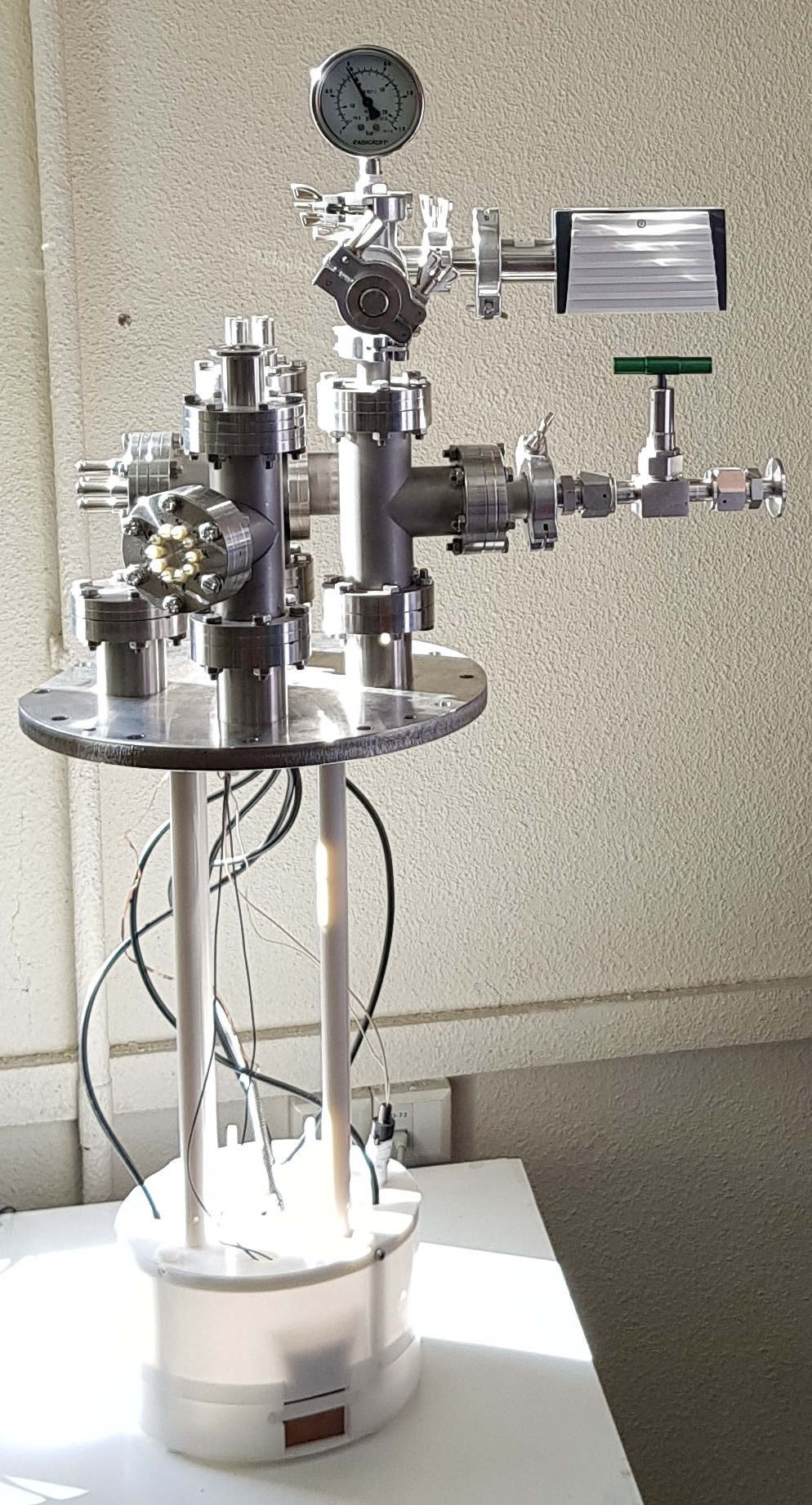}
\includegraphics[width=0.33\textwidth]{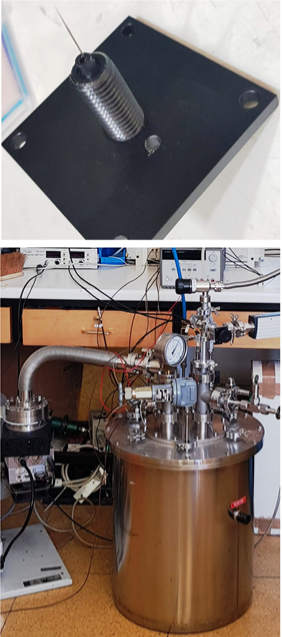}
\caption[Design and photo of the ARION internal structure]{Picture of the ARION internal structure assembly (left), tungsten needle (top-right) and cryostat (bottom-right). }
\label{setup}
\end{figure}  
\begin{paracol}{2}
\switchcolumn

The ARION experiment~(Fig.~\ref{setup} left) consists of a 1~L dual-phase argon chamber with a drift region and  charge readout. The anode is a tungsten needle with a 0.5~mm diameter, 3.8~cm height and 1~$\mu$m tip radius, located in the gas region on a Teflon holder supported by a conductive plane (Fig.~\ref{setup} top-right). The needle is used to produce a sizeable positive current, which is injected into the drift region. The cathode is a mesh of horizontal stainless steel wires with an area of 5$\times$5~cm$^{2}$. Two stainless steel field-shaping rings are placed between the anode and cathode with the goal of ensuring the uniformity of the electric field in the drift region. The bottom shaping ring is separated by 2~cm from the cathode and the top ring. The distance between the anode and cathode is adjusted mechanically, with a maximum of 7~cm. Four bars connect the support structure (Ertalyte~\cite{ERTALYTE} with a Teflon envelope) to a stainless steel vacuum flange. The chamber is enclosed in a cryostat~(Fig.~\ref{setup} bottom-right) equipped with temperature and liquid-level sensors. Five connections are used to power independently the anode, cathode, two shaping rings, and conductive plane around the needle. A vacuum of 10$^{-4}$ mbar is reached before filling the detector with argon gas at room temperature. During operation the pressure inside the cryostat is kept under 1.5 bar by a spring relief valve. 

The liquid-level inside the chamber is measured by two custom capacitance-based sensors installed vertically with the bottom part at 0.5~cm from the cathode. The first one is made of two concentric copper cylinders with a length of 12~cm. The second one is smaller, but more precise, than the first and consists of a series of parallel planes totaling 7~cm in height. The capacitance variation is used to estimate the LAr level with uncertainties of 2~mm (17\%) and 0.35~mm (5\%), respectively. The sensors are calibrated by measuring the capacitance both in air and fully covered with LN$_{2}$. The corresponding capacitance in LAr, $C_{LAr}$,  is estimated from the following expression: 

\begin{equation}\label{eq:6.23}
   C_{LAr} = \frac{\epsilon_{LAr}}{\epsilon_{LN2}} C_{LN2} ,  
\end{equation}

\noindent 
where $\epsilon_{LAr}$ and $\epsilon_{LN2}$ are the dielectric constants of LAr (1.5) and LN$_{2}$ (1.4), and $C_{LN2}$ is the capacitance measured in LN$_{2}$. The liquid-level in the detector is estimated by combining both sensors, permitting control of the argon filling rate, which is approximately 0.02~L/min. The temperature is measured with two PT100 resistors at the bottom of the level sensors.

In the experiment, the voltage of the anode is set to a maximum of 5~kV. The electrons created near the needle by the Townsend avalanche are directly collected at the plane around the needle and the positive ions drift towards the wires at the cathode along the field lines. The currents at the anode, cathode, and shaping rings, are measured using an 8-channel digital picoammeter with a maximum range of $\pm$130~nA per channel. Data-taking software based on LabVIEW~\cite{LabView} is employed to record the measured current and its standard deviation with a maximum sampling rate of 1~Hz.

One of the fundamental aspects of the experiment is to precisely tune the voltage settings of the different elements of the detector so that the complete charge produced at the anode is collected in the cathode. A thorough simulation of the setup is done with the physics simulation software COMSOL~\cite{COMSOL}, which is used to perform a finite element analysis of the electric field inside the chamber using the detector geometry and the dielectric constants. The dielectric constant used for liquid argon is 1.5. This simulation is used to study the behaviour of the field lines produced by the needle in gaseous argon under different voltage settings and temperature conditions. 

In the simulation, the voltage configuration is set to 3~kV at the anode, -3~kV at the cathode, -1~kV at the top shaping ring (T-SR), and -1.5~kV at the bottom shaping ring (B-SR). The plane (P) around the needle is set to 1~kV. The argon gas dielectric constants at 293~K (1.0005) and 98~K (1.0016) are alike, and the electric field behaviour is similar in both cases. 

The electric field map obtained from the simulation, using a similar voltage configuration as in the actual experiment, is depicted in~Fig.~\ref{sim_liquid}. The liquid argon level is set between the top and bottom shaping rings. In this study, temperatures of 85~K and 98~K are used for liquid and gas argon, respectively. The temperature gradient is assumed to be small in the charge transport region. We observe that, with the voltage settings selected for the experiment, the electric field is uniform in the drift region and most of the field lines starting at the anode terminate at the wires of the cathode. 

\begin{figure}[!t]
\includegraphics[width=0.72\textwidth]{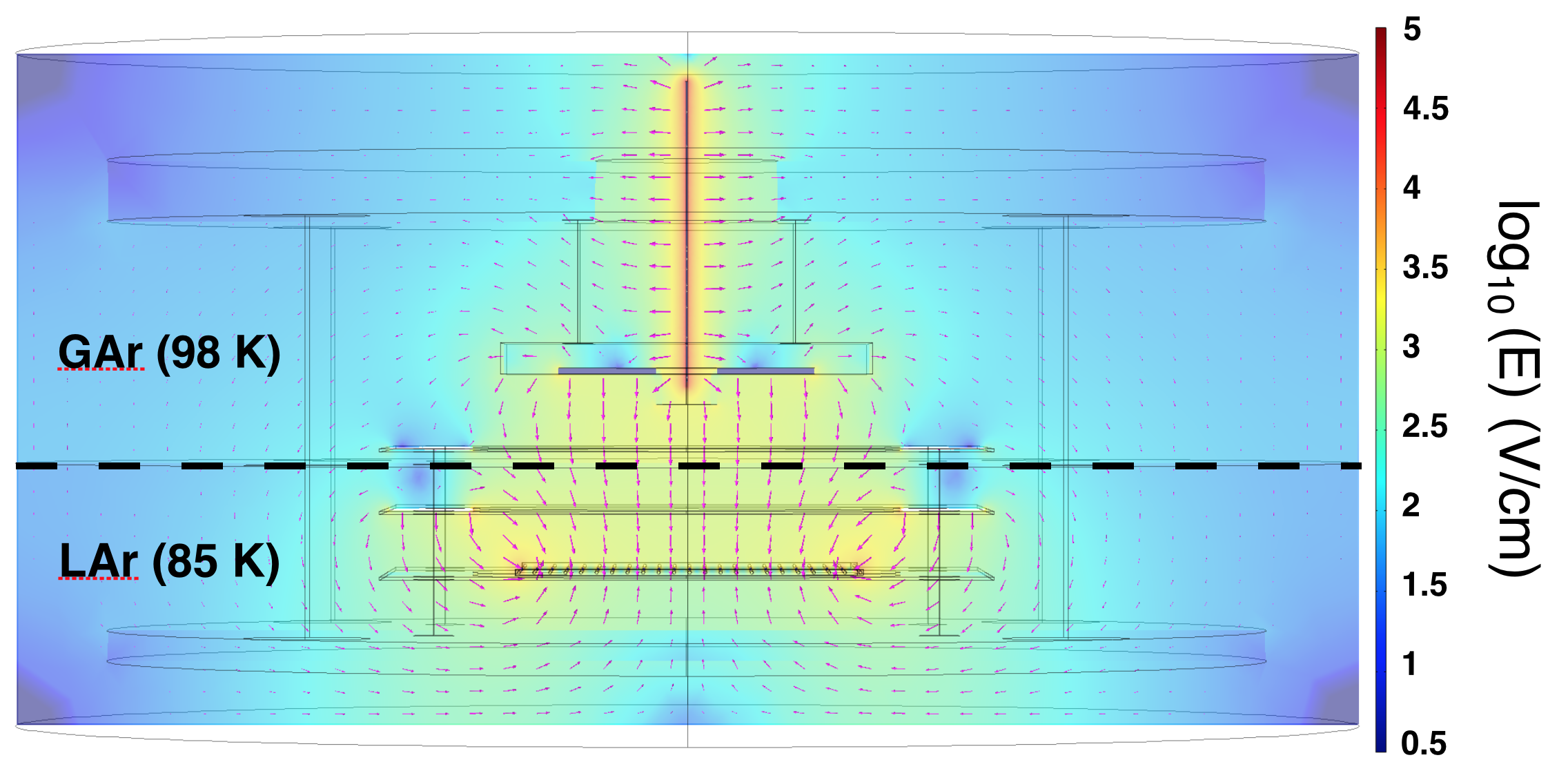}
\caption[Simulation of the electric field in the detector filled with liquid argon]{Calculation of the field lines in the detector with a finite element analysis. The level of the liquid argon is represented by the dashed line. The intensity of the field is represented in logarithmic scale. GAr denotes the argon gas volume.}
\label{sim_liquid}
\end{figure}  

\section{Study of the ion feedback from gas to liquid}

We use the ARION experiment to test different voltage configurations to maximize the charge transfer between the needle (anode) and the wire plane (cathode). 

The optimal voltage configuration used for the measurements in this section is as follows: 2 kV on the anode, -3 kV on the cathode, 0.5 kV on the plane around the needle, and 0.7 kV and 0 kV on the T-SR and B-SR, respectively.  The currents are measured in the setup parts expected to collect or produce a charge, utilizing a digital picoammeter. During the run, the pressure was raised from 1.1 to 1.5~bar in discrete steps, while the argon gas temperature was kept fixed at 293~K. Fig.~\ref{collection_hot}-left depicts the time evolution of the registered currents, along with the leakage current, defined as the sum of all the measured currents.

The avalanche at the needle is relatively stable with a measured current of approximately $-100$~nA. The ions are drifted to the cathode and collected with an efficiency of about 98\%, defined as the ratio of the current produced by the anode to that collected. The current detected in the plane around the needle and the two shaping rings is lower than 2~nA. The current produced at the anode decreases when the gas pressure increases. The leakage current remains stable to within 4\%. This result demonstrates that the experimental setup is capable of generating a sizeable, stable current between the needle and grid, allowing a study of the space charge and the dynamics of positive ions.

The detector is filled with argon gas from a Standard AirLiquide ALPHAGAZ\texttrademark-1 bottle, with an approximate purity of 10~ppm. 
We investigated the impact of argon gas purity on our measurements by comparing different runs with the argon purposely contaminated with defined amounts of air.

Fig.~\ref{collection_hot}-right displays the currents at the anode and cathode for different contamination levels at a constant pressure of 1.2~bar, with voltage configuration as described above. The current decreases by less than 4\% over the impurities range [10$^{-1}$, 10$^{4}$]~ppm. Above 10$^{5}$~ppm, the ion current produced by the Townsend avalanche is highly suppressed due to the higher ionisation energy of the air as compared to argon. These results prove that the impact of gas impurities on the ion drift is negligible.

\end{paracol}

\begin{figure}[!t]
\centering
\includegraphics[width=0.48\textwidth]{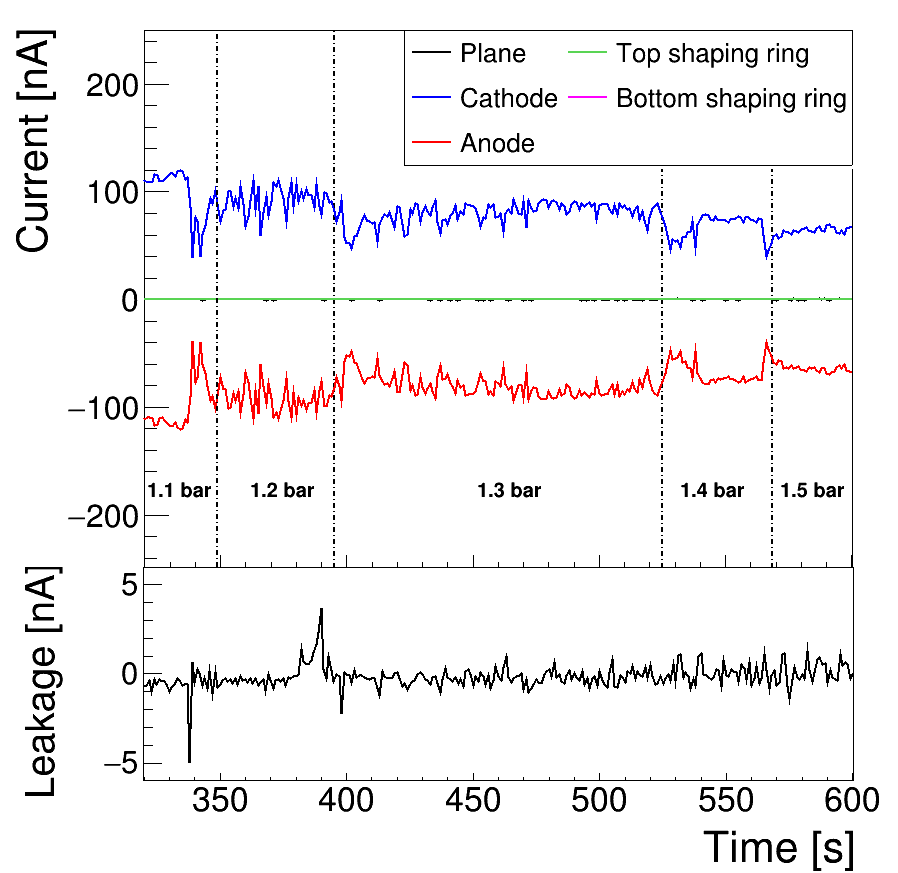}
\includegraphics[width=0.49\textwidth]{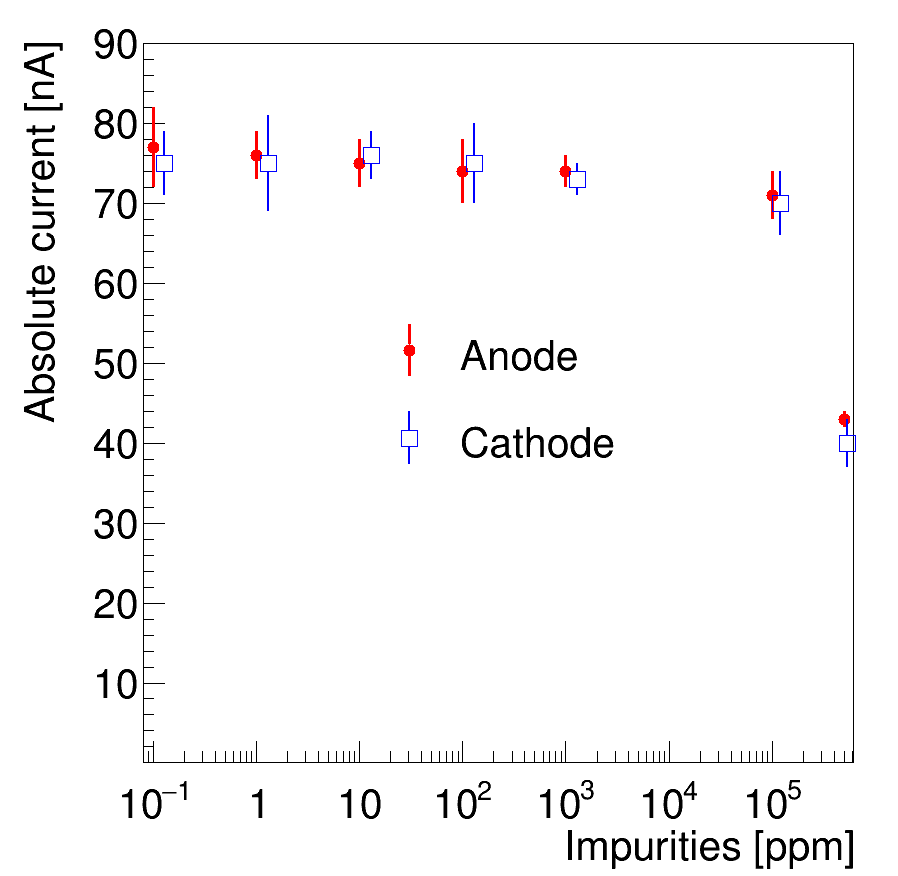}
\caption[Evolution of currents measured in the plane, cathode, anode, T-SR and B-SR]{(Left) Time evolution of currents measured at the plane (black), cathode (blue), anode (red), T-SR (green) and B-SR (magenta) with argon gas at 293~K.  The leakage current is shown in the bottom panel. (Right) Mean value of the current generated at the cathode (blue squares) and anode (red circles) for different initial contamination levels, at a pressure of 1.2~bar and voltage configuration as described in the main text.}
\label{collection_hot}
\end{figure}
\begin{figure}[!t]
\centering
\includegraphics[width=0.49\textwidth]{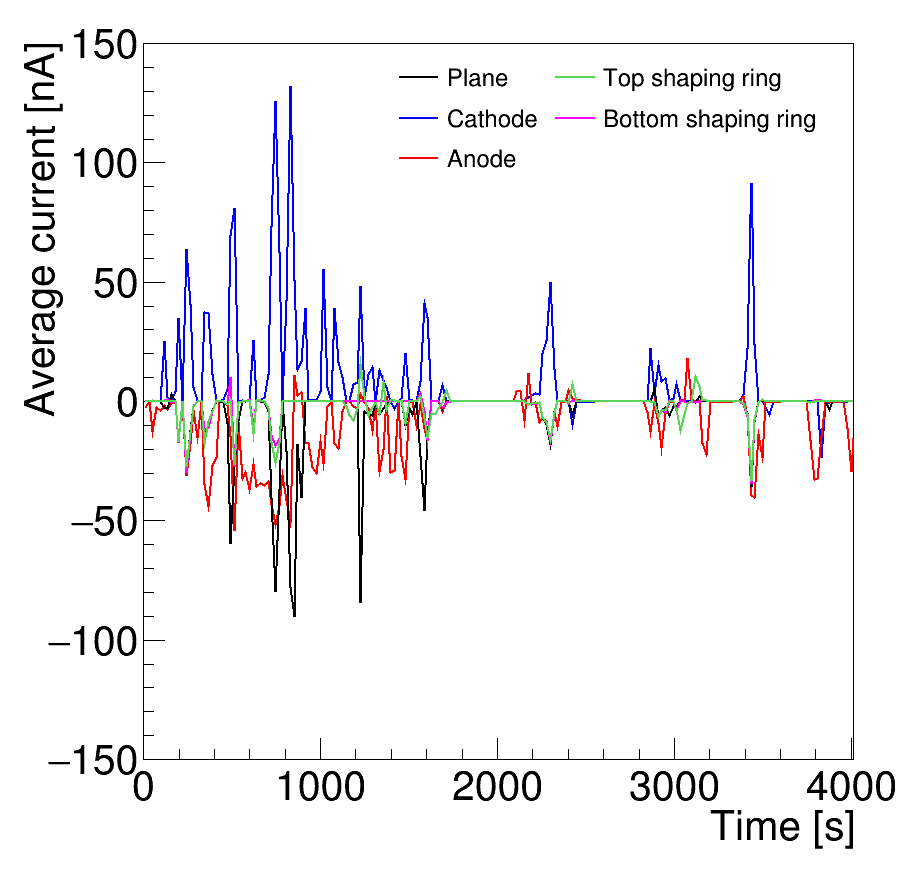}
\includegraphics[width=0.49\textwidth]{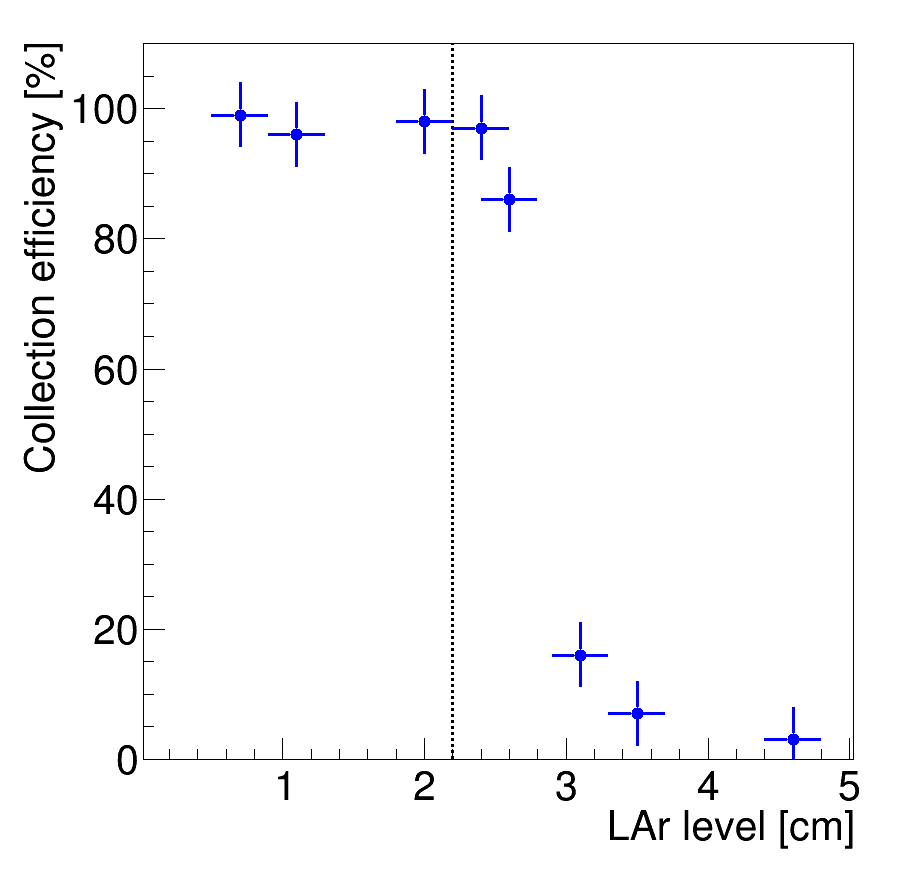}
\caption[Collection efficiency with respect to the LAr level]{(Left) Time evolution of currents measured with the cathode covered with 0.5~cm of liquid argon. Only a pulsed behaviour is observed with no stable current. (Right) Collection efficiency with respect to the LAr level, as measured from the bottom of the cryostat. The position of the cathode (2.2~cm) is indicated by the dashed line. }
\label{liquid_argon}
\end{figure}

  
\begin{paracol}{2}
\switchcolumn

A new set of measurements is performed with the detector filled with gas at 98~K, with the same voltage settings used at room temperature. The measured currents are plotted in Fig.~\ref{collection_cold}. Under these conditions, the magnitude of the ion current produced at the anode is smaller than that measured at room temperature (Fig.~\ref{collection_hot}-left). Since the voltage necessary to produce a continuous current is related to the density of the medium (pressure and temperature), this reduction in current can be explained by the increase in density of the argon gas, which is a factor of 3 higher at cryogenic than at room temperature. In this case, the collection efficiency is close to 98\%, comparable to that measured at 293~K. Our measurements demonstrate that it is possible to produce a stable ion stream between the anode and cathode with a collection efficiency close to 100\%, both at room and cryogenic temperatures.

\begin{figure}[!t]
\includegraphics[width=0.57\textwidth]{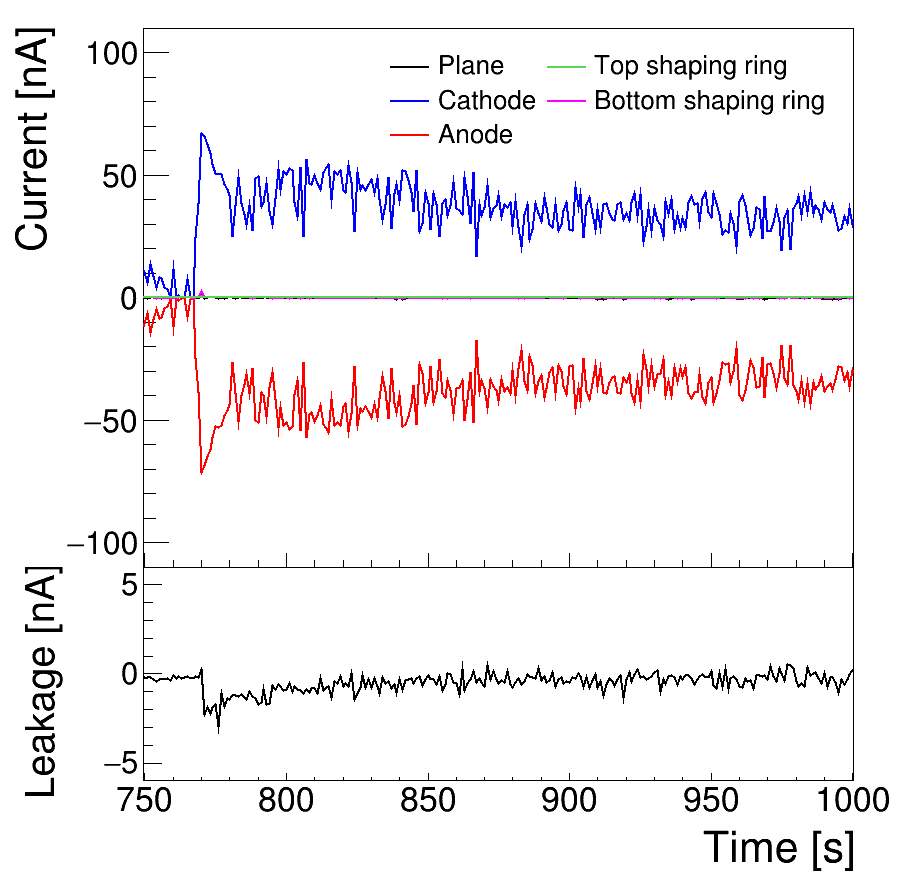}
\caption[]{Time evolution of currents measured in the plane (black), cathode (blue), anode (red), T-SR (green) and B-SR (magenta) with argon gas at 98~K. The leakage current is shown at the bottom panel.}
\label{collection_cold}
\end{figure}

To study ion feedback from the gas to liquid phase, data was collected with the chamber filled with liquid argon up to the top shaping ring. The time dependence of the currents is shown in Fig.~\ref{liquid_argon}-left. With the cathode covered by 0.5~cm of liquid argon, stable charge transfer between the anode and cathode was not observed, despite applying a voltage to the needle up to 5~kV, much higher than that previously used in gas. Instead, only current peaks were detected, characterised by quickly quenched pulses, indicating a distortion of the electric field between the needle and grid. We attribute this effect to the accumulation of positive charges in the LAr, a consequence of the low mobility of ions in the liquid. To the best of our knowledge, this is the first time space charge effects are observed and studied with a small liquid argon detector.

The collection efficiency is depicted in Fig.~\ref{liquid_argon}-right for different LAr levels, as measured from the bottom of the chamber. Independent measurements from the two liquid-level sensors are in good agreement during the initial filling. However, the high precision parallel-plane level sensor did not give reliable data for the remainder of the run. The surface of the gas-liquid interface is not perfectly flat, as bubbling in liquid argon induces roughness in the liquid-level. The sizes of these irregularities are smaller than the sensitivity of the cylindrical level meter (0.2~cm). Thus, the uncertainty on the liquid level is giving by the uncertainty of this sensor.

Similar to the results of the previous tests in cold/warm gas, the collection efficiency is close to 100\%  with the cathode not covered by LAr, while it decreases steeply with the cathode in liquid. A sizeable ion current is measured with the cathode completely submerged in LAr, proving experimentally, for the first time, the possibility for the ions produced in gas to enter the liquid, contrary to the assumption that the ions cannot enter the liquid because of the electrostatic potential problem~\cite{Bueno:2007um} and only accumulate on the surface. This result is in agreement with the model proposed in~\cite{Romero:2016tla} for the dynamic of the ions at the  gas-liquid interface.

\end{paracol}

\begin{figure}[!t]
\centering
\includegraphics[width=0.49\textwidth]{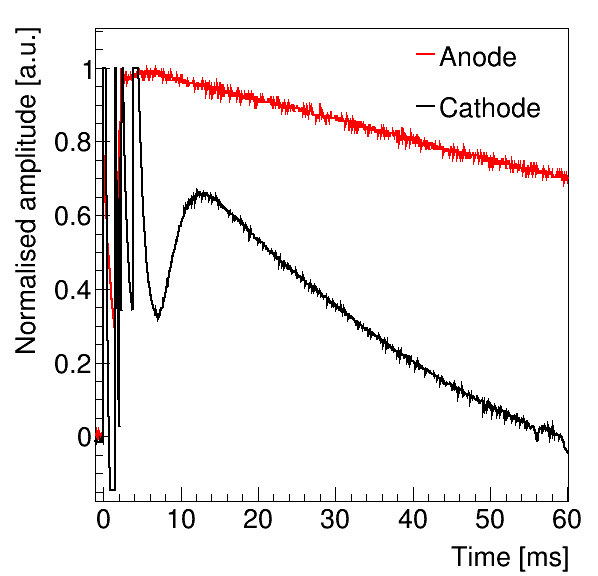}
\includegraphics[width=0.49\textwidth]{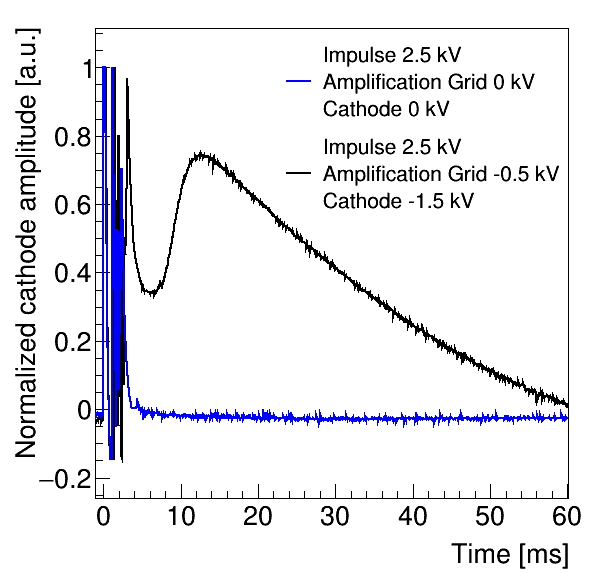}
\caption[]{(Left) Average signal introduced to the anode (red) and detected at the cathode (black). The anode is fed with a 2.5~kV pulse, the amplification grid is set to -1~kV, and the cathode to -2~kV. The signals are normalised to the maximum height of each pulse for ease of visualisation. (Right) Average signal detected at the cathode for a drift field of 180~V/cm (black) and at null field (blue). }
\label{signal_anode}
\end{figure}

\begin{paracol}{2}
\switchcolumn

\section{Test with ion pulses and preliminary measurement of the ion mobility in gas argon}

Minor changes are implemented in the experimental setup to measure the ion drift velocity in gas and liquid argon under different field conditions. This section presents the commissioning of the detector with argon gas at room temperature.

A constant voltage near ground is applied to the anode, while a negative voltage is applied to the cathode. Positive high-voltage (HV) pulses are generated using a custom-made electrical circuit~\cite{Lopez:2021} connected to the anode via a capacitor. This configuration produces a constant electric field in the drift region and allows ion bunches to be generated in a controlled manner each time a high voltage pulse is triggered. A 1~MHz oscilloscope is used to record the HV pulse (attenuatesd by a factor 1000) and the signal collected at the cathode.

Under the influence of a constant electric field, the ion cloud produced by the HV pulse drifts from the needle towards the cathode. For an electric field $E_d$, the mobility of the ions is given by:

\begin{equation}
 \mu_{i} = \frac{v_d}{E_d} = \frac{D_d}{t_d \cdot E_d} ,
\end{equation}

\noindent
where $v_d$ is the ion drift velocity, $D_d =40 \pm 0.1$~mm is the distance between anode and cathode and $t_d$ is the time necessary for the ion cloud to travel this distance. The signal from the cathode is not produced sharply at the arrival time of the ions, rather it is induced as the packet travels along its path according to the Shockley-Ramo theorem~\cite{Shockley,Ramo}:

\begin{equation}
 i = qvE_{r},
\label{eq-S-R}
\end{equation}

\noindent
where $i$ is the instantaneous electric current induced on the cathode, $q$ is the value of the charge packet, $v$ is the velocity of the packet, and $E_{r}$ is the component of the reduced field along the direction of the packet velocity.

According to Equation~\ref{eq-S-R}, ions moving at a low speed induce a tiny signal on the cathode. This signal is produced throughout the transit time of the ions in a constant electric field, making it difficult to measure the precise arrival time of the ions. This issue can be improved by replacing the wire plane with a double coplanar grid, in which one of the two grids serves as the cathode, $K$, and is imbricated on a second grid, known as the Frisch grid~\cite{Frisch:1944}, $G$, which works as an induction plane and provides a high electric field. Both grids are at different potentials, with a higher voltage on the cathode ($V_{K}$) than on the amplification grid ($V_{G}$). Assuming the needle is at ground potential, the drift voltage is $V_d = (V_{K}+V_{G})/2$ and the amplification voltage $V_A = V_{K}-V_{G}$. The significantly higher potential of $V_{K}$ compared to $V_{G}$ makes the $G$ grid transparent to drifting ions, with the charge collected only on the cathode.

The average signal over 10 pulses with argon gas at room temperature is shown in Fig.~\ref{signal_anode}-left with argon gas at room temperature at 1.4~bar. The anode is fed with a 2.5~kV pulse, the amplification grid is set to -1~kV, and the cathode to -2~kV. The anode signal rises quickly and discharges with a time constant of several tens of milliseconds. A correlated signal is measured at the cathode, in which two regions are differentiated. The first region contains several peaks with a time difference to the anode maximum peak time, $t_d$,  shorter than 8~ms. A possible explanation for these peaks is the oscillation in the cathode of the electromagnetic signal produced by the avalanche on the needle and picked up by the wires. The second region is formed by a single bump with $t_d$ larger than 10~ms.

Fig.~\ref{signal_anode}-right illustrates the average pulse at the cathode with $V_d = 1$~kV and the cathode at ground ($V_{K} = 0$). When the drift field is removed, the slow component of the signal disappears and only fast pulses associated with electron induction signals are detected. Furthermore, the time position of the bump observed in the signal at the cathode changes when the values of the electric field or pressure are modified. We therefore conclude that the observed bump corresponds to the positively charged cloud generated at the needle.

\end{paracol}

\begin{paracol}{2}
\switchcolumn

Fig.~\ref{ion_velocity} depicts the ion mobility in argon gas in the range [1.1-1.4]~bar, calculated averaging over 10 pulses. The analog manometer introduces an uncertainty of 0.2~bar on the pressure measurement. Ion mobility values in the range [1.8-1.7]~$\text{cm}^2\,\text{V}^{-1}\text{s}^{-1}$ are measured for drift fieldThe abstract has been rewrittens between 120 and 260~$\text{V}\,\text{cm}^{-1}\text{bar}^{-1}$. The statistical uncertainty is considered for each point. These values are similar to the ones obtained previously in~\cite{Deisting:2018vtx,Madson:1967}, confirming the capability of the ARION experiment to measure the ion drift velocity. The variation of the ion mobility with the drift field is described in two different models. The first one assumes a constant relation, as expected for low electric fields~\cite{Deisting:2018vtx,Neves:2010}, and the second model considers a dependence of the form  $\mu_{i} \approx 1/\sqrt{E_d}$, as suggested for high electric fields in~\cite{Blum:2008nqe}. Fig.~\ref{ion_velocity} shows that the data are best described by the low electric field model.

\begin{figure}[!t]
\includegraphics[width=0.57\textwidth]{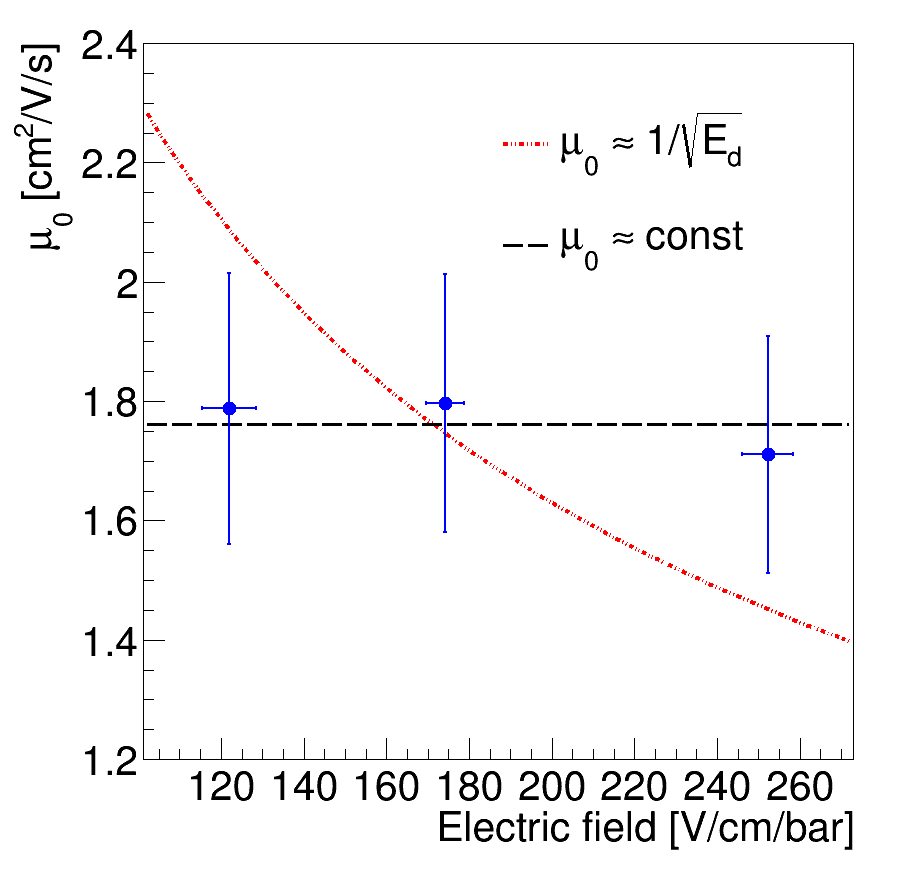}
\caption[]{Ion mobility in argon gas at room temperature with respect to the intensity of the drift field considering two different dependencies with the electric field.}
\label{ion_velocity}
\end{figure}

An electric field is considered low if the energy gained by an ion in the field is much smaller than the thermal energy. Considering pure argon this is equivalent to~\cite{Simon:2017}:   

\begin{equation}
   E_d \ll kT/2\ell e,
\end{equation}

\noindent
where $e$ is the electron charge, $k$ the Boltzmann constant, $T$ the temperature of the gas (293~K), and $\ell$ the mean free path between collisions, with a value of 49~nm in pure argon at 1.4~bar~\cite{Jousten:2018}. Using these values, we obtain $E_d \ll $2~kV/cm, thus the low-field approximation is valid in the range of operating electric fields studied in this paper. We have therefore demonstrated the ability of the ARION setup to accurately detect ion pulses and measure the drift velocity of the ion packet.

\section{Conclusions} 

The significantly lower drift velocity of positive ions, compared to that of electrons, can induce an sizeable positive charge density in a LAr detector. A resulting volume charge would modify the drift field and quench the negative charge signal. In the case of dual-phase detectors with charge amplification, the positive charge density can be further increased by the ion feedback from the gas to the liquid phase. This effect may be particularly relevant for the next generation of massive argon detectors for dark matter searches and neutrino studies, establishing an intrinsic limit to the technology and reducing the maximum electron drift distance attainable even in the presence of infinite liquid purity.  

With the aim of studying the dynamics of positive ions in liquid argon, we have constructed a small detector (1~L) capable of generating, drifting and collecting sizeable positive charge currents. Our setup has been successfully operated to obtain the first evidence of ion feedback from the gas to liquid phase using a continuous ion current. The detector has been successfully run with argon gas at room temperature using ion pulses generated in a controlled manner. Under these conditions, the ion mobility has been measured to be in the range [1.8-1.7]~$\text{cm}^2\,\text{V}^{-1}\text{s}^{-1}$ for drift fields between 120 and 260~$\text{V}\,\text{cm}^{-1}\text{bar}^{-1}$. Additionally, we have shown that the ion mobility does not depend on the electric drift field. Additional measurements with liquid argon will be reported in a future publication.

\vspace{6pt} 

\authorcontributions{Conceptualization, L.R., R.S. and T.L.; methodology, L.R. and E.S.G.; formal analysis, S.Q., M.L., S.L. and E.S.G.; investigation, V.P., R.L.M. and J.M.C.; writing---original draft preparation, R.S. and E.S.G.; writing---review and editing, M.L. and P.G.A. All authors have read and agreed to the published version of the manuscript.}

\funding{The research has been funded by the Spanish Ministry of Economy and Competitiveness (MINECO) through the grant PID2019-109374GB-I00 (MICINN). The authors were also supported by the ``Unidad de Excelencia Mar\'{i}a de Maeztu: CIEMAT - F\'{i}sica de part\'{i}culas"  through the grant MDM-2015-0509. T.~Lux is funded by the Spanish Ministerio de Economía y Competitividad (SEIDI - MINECO) under Grants No. PID2019-107564GB-I00 and SEV-2016-0588. IFAE is partially funded by the CERCA program of the Generalitat de Catalunya. M.~Leyton acknowledges funding from the Marie Skłodowska-Curie Fellowship program, under grant 665919 (EU, H2020-MSCA- COFUND-2014).}

\conflictsofinterest{The authors declare no conflict of interest.} 



\end{paracol}
\reftitle{References}




\end{document}